# Localized Photonic jets from flat 3D dielectric cuboids in the reflection mode

I.V. Minin,[1] O. V. Minin,[1] V. Pacheco-Peña[2*], M. Beruete[2]

[1]*Siberian State Academy of Geodesy, Plahotnogo 10, Novosibirsk, 630108, Russia*
[2]*Antennas Group - TERALAB, Universidad Pública de Navarra, Campus Arrosadía, 31006 Pamplona, Spain*
*\*Corresponding author: miguel.beruete@unavarra.es*

A photonic jet (terajet at THz frequencies) commonly denotes aspecific spatially localized region in the near-field at the front side of a dielectric particle with diameter comparable with wavelength illuminated with a plane wave from its backside (i.e., the jet emerges from the shadow surface of a dielectric particle). In thispaper the formation of photonic is demonstrated using the recently proposed 3D dielectric cuboids working in "reflection" mode when the specific spatially localized region is localized towards the direction of incidence wavefront. The results of simulations based on Finite Integration Technique are discussed. All dimensions are given in wavelength units so that all results can be scaled to any frequency of interest including optical frequencies, simplifying the fabrication process compared with spherical dielectrics.. The results here presented may be of interest for novel applications including microscopy techniques and sensors. © 2014 Optical Society of America

It is known that several applications, such as those in microscopy, require the focusing of an incoming plane wave into a spot (focus) with a spatial resolutionsmaller than the incident wavelength. However, the limit imposed by the diffraction of electromagnetic (EM) waves should be addressed in order to generate a subwavelength focal spot [1]. This limit in spatial resolution has been extensively studied recently leading to different techniques [2–7].

Several years ago, a way to overcome the diffraction limit using microscaled cylindrical (2D) and spherical (3D) dielectrics under plane wave illumination was reported [8–18], demonstrating the capability to produce photonic nanojets (PNJ) with a resolution of $\lambda_0/3$ (where $\lambda_0$ is the operation wavelength). Also, the capability to produce jets at terahertz (THz) and sub-THz frequencies (so called terajets) has been recently proposed and demonstrated experimentally by using 2D and 3D dielectric cuboids [19,20]. It has been demonstrated that such structures have similar performance compared with the spherical dielectrics, demonstrating that subwavelength spatial resolution can be reached when a contrast of refractive index between the 3D cuboid and the background medium is less than 2:1. It has been reported that terajets formation based on dielectric cuboids is possible not only at the fundamental harmonic, but also at other frequency harmonics for normal and oblique incidence [20]. Such structure may be considered as a planarjet lens emerging from the shadow surface of its output face.

Various practical applicationsrequire the creation of different types of photonic jets orstreams with specific characteristics and properties. To date, the formation of photonicjets and their applications have been developed using dielectric particles in the transmission regime. i.e., the dielectric particles are illuminated with a plane wave from their back and the jet is formed just at their front (along the propagation direction of the incident radiation).

In this paper, the capability to produce photonic jets is evaluated using 3D dielectric cuboids in the reflection regime. First, the capability to produce photonic jets is numerically studied using dielectric cuboids with different dimensions along the optical $z$-axis. The focusing properties are evaluated in terms of focal length (FL), full-width at half-maximum along both transversal $x$- and $y$-axis (FWHM$_x$ and FWHM$_y$, respectively) at each focal length and the ellipticity (defined as the ratio between both transversal resolutions FWHM$_x$/FWHM$_y$).Simulation results demonstrate that the photonic jet is located inside the dielectric cuboid when its dimension along $z$ is greater than $0.4\lambda_0$'at the operation wavelength. Moreover, the optimum thickness in terms of ellipticity is $\sim 0.33\lambda_0$' with results in an ellipticity$\sim 1.07$; i.e., an almost spherical focus is achieved at the focal position. The multifrequency response is also evaluated at the first harmonic ($\lambda_1'=0.5\lambda_0'$) demonstrating that subwavelength focusing is also reachable working in the reflection regime at frequency harmonics. Finally, the photonic jet performance is evaluated at the fundamental frequency under oblique incidence by rotating the 3D dielectric cuboid from $0°$ to $15°$ demonstrating that the enhancement at the focal position is not deteriorated with respect to the value obtained under normal incidence.

To begin with, it is known that the effects of photonic jets cannot be predicted on the basis of geometrical optics or scalar diffraction theory.Therefore it is essential to study the propagation of electromagnetic waves through such elements using Maxwell's equations. To evaluate the photonic jet performance, the numerical results in this work are carried out using the transient solver of the commercial software CST Microwave Studio™ with the same boundary conditions and mesh as in [19,20]. Moreover, since the 3D dielectric cuboids can be scaled at different frequency ranges, here both wavelength and frequency will be normalized as in [20] as $\lambda' = \lambda/L$ and $f' = fL/c$; where the unprimed and primed ones are the physical and normalized values, respectively, $c$ is the velocity of light in vacuum and $L$ is the dimension of the cuboid along both transversal $x$- and $y$-axes.

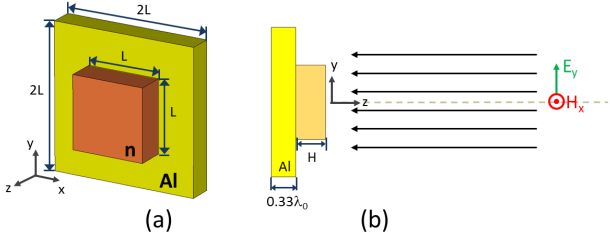

Fig. 1. Schematic representation of the proposed 3D dielectric cuboid working in the reflection mode placed at the front of a metallic plate of aluminum with dimensions $2L \times 2L \times 0.33L$. The 3D dielectric cuboid of refractive index $n = 1.46$ has transversal dimensions $L = \lambda_0'$ along both $x$- and $y$- axis and thickness $H$: perspective view (a) and lateral view (b). The whole structure is immersed in vacuum ($n_0 = 1$).

The 3D dielectric cuboid used in this letter in the reflection regime is schematically shown in Fig.1 (a). The cuboid, with a refractive index of $n = 1.46$, has lateral dimensions $L = \lambda_0'$ and thickness $H$ and is placed at the front side of a metallic plate of aluminum (with a conductivity $\sigma_{Al} = 3.56 \times 10^7$ S/m) of dimensions $2L \times 2L \times 0.33L$ along $x$-, $y$- and $z$- axis, respectively. A vertically polarized ($E_y$) plane wave is used as a source illuminating the cuboid from its front side; i.e., with a propagation along the negative $z$- axis [see Fig. 1(b)]. To evaluate the photonic jets produced in reflection mode, the whole structure is considered to be immersed in vacuum ($n_0 = 1$).

First, the focusing performance of the photonic jets is evaluated at the fundamental frequency ($f_0' = 1$, $\lambda_0' = 1$) using several 3D dielectric cuboids with the same transversal dimensions (along $x$- and $y$- axis) and changing the dimension along the optical axis ($z$- axis). Numerical results of the power distribution on the $xz$ plane/$H$-plane for different values of the cuboid thickness $H$ from $0.2\lambda_0'$ to $0.8\lambda_0'$ with a step of $0.2\lambda_0'$ are shown in the left column of Fig. 2. It can be observed that focusing is produced for all the geometries; however, for values of $H=0.6\lambda_0'$ and $H=0.8\lambda_0'$ the focus is inside the 3D dielectric cuboid. By decreasing the axial dimension below $0.4\lambda_0'$, it is shown that the photonic jet is moved toward the cuboid and it is close to the output surface when $H=0.4\lambda_0'$ and completely outside the cuboid for $H=0.2\lambda_0'$. In order to better compare these results, the normalized power distribution along the optical $z$-axis is shown on the left column of Fig. 2 for each value of $H$ (the output surface of the 3D dielectric cuboid is plotted an orange dotted line in all cases as a reference and the power distribution without the presence of the cuboid is plotted as a gray line). From these results, the optimum value of H should be $H<0.4\lambda_0'$, in order to produce the photonic jet just at the surface of the cuboid for the fundamental frequency. Note that, this dimension is optimal for the refractive index here used and this should be changed for different values of $n$, for example, if $H$ is selected to be $0.8\lambda_0'$ a value of $n=1.2$ (not shown here) should be used for the cuboid in order to obtain the photonic jets just at the surface of the dielectric. Based on this, with the aim of studying the features of the photonic jets produced at the output surface of the cuboid, several structures were evaluated for values of $H$ from $0.2\lambda_0'$ to $0.4\lambda_0'$ with a step of $0.7\lambda_0'$ at the fundamental frequency ($f_0' = 1$, $\lambda_0' = 1$) and the first frequency harmonic ($f_1' = 2$, $\lambda_1' = 0.5$). The focusing properties are summarized in Table 1 and its performance is evaluated in terms of the focal length (FL), power enhancement (defined as the ratio of the power distribution at the FL with and without the presence of the dielectric cuboids) and *photonic jet exploration range* ($\Delta l_z$), defined as the distance from the FL at which the intensity enhancement has decayed to half its maximum value. Note that in this case the maximum value is selected to be at the FL and not at the surface of the cuboid as in [19]. This is due to the fact that the focus may be outside of the cuboid and not just at the surface for different values of H, as it has been explained before. It can be observed that the best performance at the fundamental frequency is obtained for a value of $H=0.33\lambda_0'$, where the photonic jet is obtained closer to the output surface of the dielectric cuboid and a maximum enhancement of ~7.3 is achieved. On the other hand, when the first frequency harmonic is used, the best performance is achieved for a value of $H=0.4\lambda_0'$ with an enhancement of ~17. Note that, in general, the enhancement achieved with the first frequency harmonic is higher for all the values of H here evaluated with respect to those obtained at the fundamental frequency, which is in good agreement with the recently proposed cuboid working in the transmission regime [20]. Therefore, this structure may be used for both frequency harmonics where the photonic jet performance is not deteriorated.

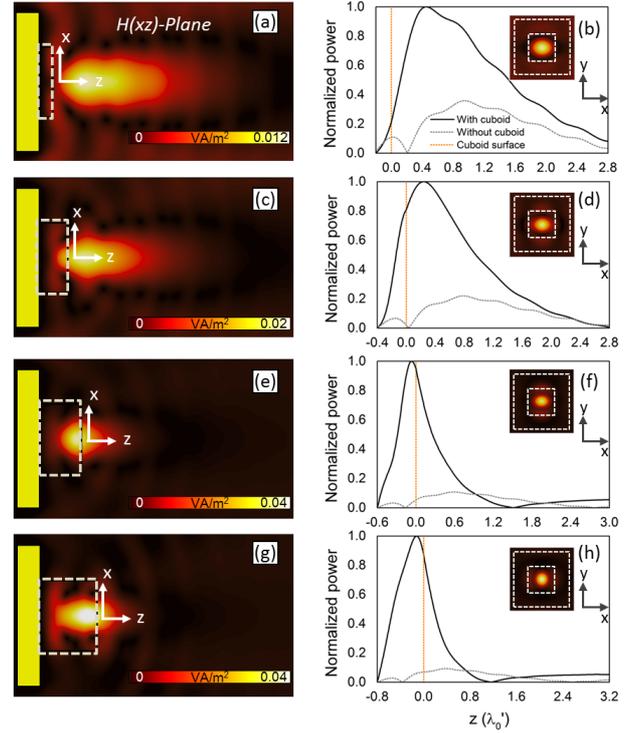

Fig. 2. Numerical simulations of the power distribution on the $xz$-plane/$H$-plane for different cuboids (left column) when the transversal dimensions are fixed with a value of $L \times L = \lambda_0' \times \lambda_0'$ along $x$- and $y$- axis, while the optical dimension along $z$- axis is changed with values of: (a)$H=0.2\lambda_0'$, (c)$H=0.4\lambda_0'$,(e)$H=0.6\lambda_0'$ and (g)$H=0.8\lambda_0'$. Numerical results of the normalized power distribution along the optical $z$-axis (right column) for different values of H:(b)$H=0.2\lambda_0'$, (d)$H=0.4\lambda_0'$,(f)$H=0.6\lambda_0'$ and (h)$H=0.8\lambda_0'$. The output surface of the 3D dielectric cuboid is plotted as orange dotted line, the power distribution without the presence of the dielectric cuboid is plotted as gray dotted line and the power distribution on the $xy$-plane at each focal length is shown as inset (the cuboid and the metal square are plotted as white dashed lines.

For the sake of completeness, numerical results of the full-width at half-maximum along both transversal $x$- and $y$- axis (FWHM$_x$ and FWHM$_y$ respectively) at the fundamental and first frequency harmonic are plotted in Fig. 3 (a-b) for the same range of $H$. Also, the ellipticity for both frequencies are plotted in Fig. 3 (c). For the case of the fundamental frequency, it is shown that again the best performance is achieved when $H=0.33\lambda_0'$ with a resolution of FWHM$_x$ = $0.44\lambda_0'$ and FWHM$_y$ = $0.41\lambda_0'$. With this design, the best ellipticity is obtained with a value of 1.073, i.e., a quasi-spherical spot is achieved. On the other hand, the best design for the first frequency harmonic is also obtained when $H=0.4\lambda_0'$ with values of FWHM$_x$ = $0.55\lambda_1'$ and FWHM$_y$ = $0.57\lambda_1'$ with an ellipticity of 0.96 which is also an almost spherical focus.

It is also interesting to evaluate the focusing performance of the 3D dielectric cuboid in reflection regime under oblique incidence. Numerical results of the the normalized power distribution on the $xz$-plane/$H$-plane are shown in Fig. 4 at the fundamental frequency ($f_0'$ = 1, $\lambda_0'$ = 1) when the whole structure is rotated from 5° to 15° with a step of 5°. The dimensions of the cuboid are $\lambda_0' \times \lambda_0' \times 0.33\lambda_0'$ along $x$-, $y$- and $z$-axis, respectively; which are the dimensions of the cuboid with the best focusing performance at this frequency, as it has been explained before. In Fig. 4, a spillover of energy around the edges of the cuboids is observed in the direction of rotation. As a result of the destructive interference, the shape of the photonic jet is distorted. Regarding the focusing enhancement the values are ~7.14, ~7.39 and ~7.4 for the rotation angles 5°, 10° and 15°, respectively, demonstrating that the photonic jet is preserved and its performance is not fully deteriorated. Also, it can be observed that the direction of the photonic jet is clearly dependent on the rotation angle and therefore the photonic jet may be rotated in the space.

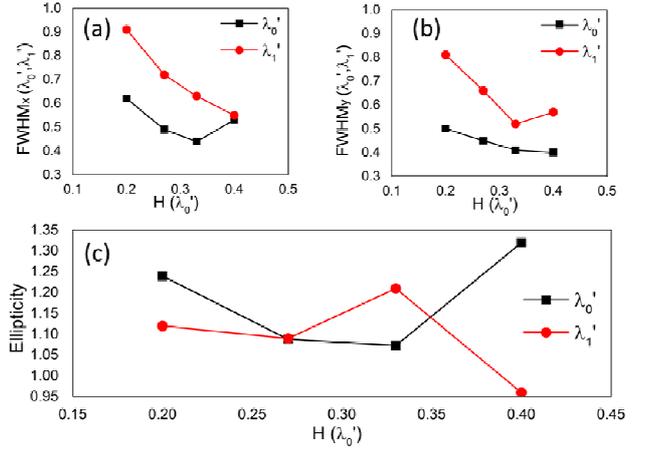

Fig. 3. Numerical results for the fundamental (black curves) and first frequency harmonic (red curves) of: (a) full-width at half-maximum along the $x$-axis, (b) full-width at half-maximum along the $y$-axis and (c) the ellipticity defined as the ratio between both spatial resolutions FWHM$_x$/FWHM$_y$.

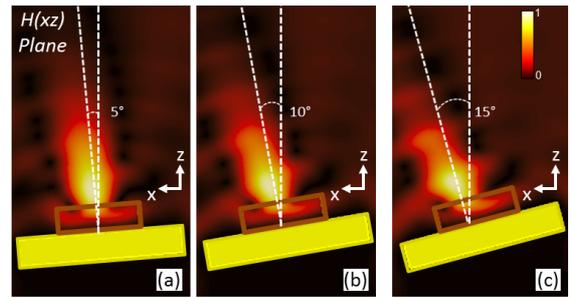

Fig. 4. Simulation results of the normalized power distribution on the $xz(H)$plane for the 3D dielectric cuboid under oblique incidence at the fundamental frequency ($\lambda_0'$ = 1) when the whole structure is rotated at (a) 5°, (b) 10° and (c) 15°.

Finally, Fig. 5 shows the normalized power distribution when the width (dimension along the transversal $x$-axis) of the cuboid is increased to reach the metal plate (i.e., the total width is $1.5\lambda_0'$) and the whole structure is rotated at 10° relative to the direction of the incident planewave. Now the area of high field concentration corresponding to the photonic jet is not perpendicular, but substantially parallel to the flat surface of the cuboid and the photonic jet is widened and narrowed along the $x$-axis and $z$-axis, respectively. Numerical results of the normalized power distribution along the transversal $x$- axis at the FL ($z=0.249\lambda_0'$) and along the optical axis at $x=-0.21\lambda_0'$ are shown in Fig. 5 (a) and (b), respectively. From these results the values of FWHM$_x$ and photonic jet exploration range are ~0.96$\lambda_0$' and 0.245$\lambda_0$', respectively. Moreover, the enhancement achieved with this configuration is ~ 5.6; i.e, the enhancement is reduced in 1.79 in comparison with the geometry shown in Fig. 4 (b).

Table 1. Numerical results of the photonic jet performance in reflection mode at the fundamental ($f_0'$=1, $\lambda_0'$=1) and first frequency harmonic ($f_0'$=2, $\lambda_0'$=0.5) using 3D dielectric cuboids with the same transversal dimensions and different values of optical dimension H.

| Frequency | H | FL[a] | Enhancement | $\Delta_z$[b] |
|---|---|---|---|---|
| $f_0'$ = 1 | $0.4\lambda_0'$ | $0.245\lambda_0'$ | ~7.25 | $0.71\lambda_0'$ |
| $f_1'$ = 2 |  | $0.874\lambda_1'$ | ~17 | $1.08\lambda_1'$ |
| $f_0'$ = 1 | $0.33\lambda_0'$ | $0.188\lambda_0'$ | ~7.3 | $0.789\lambda_0'$ |
| $f_1'$ = 2 |  | $0.749\lambda_1'$ | 10.3 | $1.32\lambda_1'$ |
| $f_0'$ = 1 | $0.27\lambda_0'$ | $0.269\lambda_0'$ | ~6.6 | $0.796\lambda_0'$ |
| $f_1'$ = 2 |  | $1.33\lambda_1'$ | ~13.6 | $1.74\lambda_1'$ |
| $f_0'$ = 1 | $0.2\lambda_0'$ | $0.45\lambda_0'$ | ~4.5 | $1.06\lambda_0'$ |
| $f_1'$ = 2 |  | $2.12\lambda_1'$ | ~7.8 | $3.1\lambda_1'$ |

[a]FL is the focal length.
[b]$\Delta_z$ is the photonic jet exploration range.

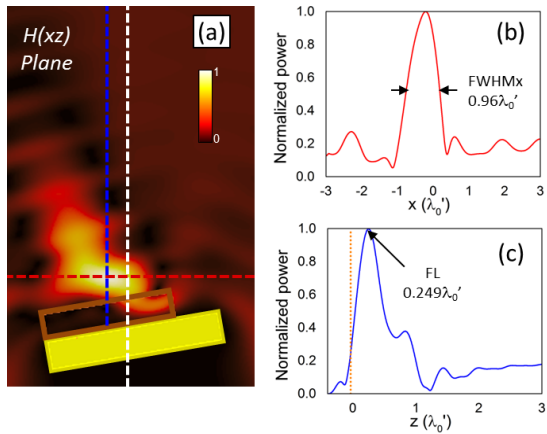

Fig. 5. (a) Normalized power distribution on the xz-plane/$H$-plane when the width of the cuboid is increased to reach the metal plate and the whole structure is rotated 10º. Normalized power distribution along the transversal $x$-axis (b) at the FL (z=0.249$\lambda_0$') and along the optical $z$-axis (c) at x=-0.21$\lambda_0$'.

In conclusion, the capability to produce photonic jets using a 3D dielectric cuboid of refractive index $n$=1.46 immersed in vacuum ($n_0$= 1) working in reflection mode has been studied. The focusing properties of the spot produced by the cuboid have been evaluated at the fundamental and first frequency harmonic under normal incidence when the dimension along $z$-axis ($H$) is changed. The photonic jet performance has been studied in terms of spatial resolution (FWHM), FL, enhancement, photonic jet exploration range and ellipticity. The results here presented demonstrate that the best performance for the fundamental and first frequency harmonic are obtained when $H$=0.33$\lambda_0$' and $H$=0.4$\lambda_0$', respectively, with a quiasi-spherical and subwavelength spot at the FL in both cases. Also, the photonic jets produced have been studied under oblique incidence by rotating the reflecting structure for angles from 5º to 15º demonstrating that a high enhancement is achievable even when the structure is not illuminated at 0º. It was also shown that we can control the space position of photonic jet up to parallel to the flat surface of the cuboid. Due to our limitations in fabrication and measurements, experimental demonstrations have not been possible yet; however we hope to solve it in the near future. The results here presented may be scaled to other frequency bands such as optical frequencies and may be applied to enhance the radiation performance of antennas, novel microscopy techniques and sensors.

This work was supported in part by the Spanish Government under contract TEC2011-28664-C02-01. V.P.-P. is sponsored by Spanish Ministerio de Educación, Cultura y Deporte under grant FPU AP-2012-3796. M.B. is sponsored by the Spanish Government via RYC-2011-08221.